\documentclass{aa}
\usepackage{graphicx}
\usepackage{color}
\usepackage{epstopdf}
\usepackage{txfonts}
\begin{document}
\title{Circumstellar water vapour in M-type AGB stars: \\Constraints from $\rm{H_2O}$(1$_{10}$--1$_{01}$) lines obtained with Odin\footnote{Based on observations with Odin, a Swedish-led satellite project funded jointly by the Swedish National Space Board (SNSB), the Canadian Space Agency (CSA), the National Technology Agency of Finland (Tekes), and Centre National dՃtudes Spatiales (CNES). The Swedish Space Corporation was the industrial prime contractor and is operating Odin.}}

\titlerunning{Circumstellar $\rm{H_2O}$ line observations with the Odin satellite}

   \author{M. Maercker
           \inst{1},
          F. L. Sch\"oier \inst{2},
	 H. Olofsson\inst{1,2},
	  P. Bergman\inst{3},
	  U. Frisk\inst{4},
	  \AA. Hjalmarson\inst{2}
	 K. Justtanont\inst{2},
	 S.Kwok\inst{5,6},
	  B.~Larsson\inst{1},
	  M. Olberg\inst{2},	 
	 \and
	 Aa. Sandqvist\inst{1}
	  }
\authorrunning{M. Maercker et al.}

\offprints{M. Maercker}

   \institute{
   	Stockholm Observatory, AlbaNova University Center, 106 91 Stockholm, Sweden\\
       	\email{maercker@astro.su.se}
    	\and
     	Onsala Space Observatory,  439 92 Onsala, Sweden
         \and
        	European Southern Observatory, Casila 19001, Santiago 19, Chile
        	\and
     	Swedish Space Corporation, PO Box 4207, 171 04 Solna, Sweden 
	\and
	Department of Physics \& Astronomy, University of Calgary, Calgary, Canada
	\and
	Department of Physics, University of Hong Kong, Hong Kong}

   \date{Received ; accepted }

 \abstract
{A detailed radiative transfer code has been previously used to model circumstellar ortho-$\rm{H_2O}$ line emission towards six M-type asymptotic giant branch stars using Infrared Space Observatory Long Wavelength Spectrometer data. Collisional and radiative excitation, including the $\rm{\nu_2=1}$ state, was considered.}
{Spectrally resolved circumstellar $\rm{H_2O}$(1$_{10}$--1$_{01}$) lines have been obtained towards three M-type AGB stars using the Odin satellite. This provides additional strong constrains on the properties of circumstellar $\rm{H_2O}$, in particular on the chemistry in the stellar atmosphere, and the photodissociation in the outer envelope.}
{Infrared Space Observatory and Odin satellite $\rm{H_2O}$ line data are used as constraints for radiative transfer models. Special consideration is taken to the spectrally resolved Odin line profiles, and the effect of excitation to the first excited vibrational states of the stretching modes  ($\rm{\nu_1=1}$ and $\rm{\nu_3=1}$) on the derived abundances is estimated. A non-local, radiative transfer code based on the accelerated lambda iteration formalism is used. A statistical analysis is performed to determine the best-fit models.}
{The $\rm{H_2O}$ abundance estimates are in agreement with previous estimates. The inclusion of the Odin data sets stronger constraints on the size of the $\rm{H_2O}$ envelope. The $\rm{H_2O}$(1$_{10}$--1$_{01}$) line profiles require a significant reduction in expansion velocity compared to the terminal gas expansion velocity determined in models of CO radio line emission, indicating that the $\rm{H_2O}$ emission lines probe a region where the wind is still being accelerated. Including the $\rm{\nu_3=1}$ state significantly lowers the estimated abundances for the low-mass-loss-rate objects. This shows the importance of detailed modelling, in particular the details of the infrared spectrum in the range 3 to 6\,$\mu$m, to estimate accurate circumstellar $\rm{H_2O}$ abundances.}
{Spectrally resolved circumstellar $\rm{H_2O}$ emission lines are important probes of the physics and chemistry  in the inner regions of circumstellar envelopes around asymptotic giant branch stars. Predictions for $\rm{H_2O}$ emission lines in the spectral range of the upcoming Herschel/HIFI mission indicate that these observations will be very important in this context.}


   \maketitle

\section{Introduction}
\label{intro}

The asymptotic giant branch (AGB) marks the final stage in the life of low- to intermediate-mass stars ($\approx0.8-10\,$M$_{\odot}$). During this stage the stars lose a significant fraction of their mass in an intense wind, forming a circumstellar envelope (CSE). After the AGB, what remains is a C/O core, which illuminates the expanding CSE for a short while, forming a planetary nebula (Habing~\cite{habing}).

The mass loss dominates the evolution of a star on the AGB, and the physical conditions and the elemental abundances dictate the environment for the chemistry. It is therefore essential to try to understand the conditions in the inner CSE of an AGB star, in order to get a handle on the mass-loss mechanism (in particular for M-type stars; Woitke~\cite{woitke}, H\"ofner \& Andersen~\cite{hofnerco}), and the molecule and dust formation (e.g. Cherchneff~\cite{cherchneff}).

The chemistry in the wind is set by the relative abundances of carbon and oxygen. `Oxygen-rich' (M-type) AGB stars have C/O ratios that are $<1$, resulting in a dominance of oxygen-bearing molecules (Olofsson~\cite{olofssonagb}, and references therein), as most of the carbon is bound in CO. Carbon stars (C-stars), on the other hand, have C/O ratios $>1$ and therefore produce mainly carbon-bearing molecules. However, molecules typical of an M-type environment have been found in reasonably high abundances in carbon-rich AGB stars and vice versa (e.g. Olofsson et al.~\cite{olofssonetal91}, Ag\'undez \& Cernicharo~\cite{agundezcernicharo}, Sch\"oier et al.~\cite{schoieretal06}), indicating a departure from thermodynamical chemical equilibrium.

Observations of water vapour in AGB stars are important, as $\rm{H_2O}$ is one of the major (and possibly the dominant) coolant in the CSEs (Truong-Bach et al.~\cite{truongbachetal}), with a large amount of far-infrared (FIR) transitions. $\rm{H_2O}$ is expected to be the most abundant trace molecule next to CO in M-type AGB stars, and as a source of circumstellar O and OH it is important for the chemistry in the envelope. Indeed, high amounts of $\rm{H_2O}$ have been found in M-type AGB stars (e.g. Justtanont et al. \cite{justtanontetal}; Maercker at al.~\cite{maerckeretal}, hereafter Paper I). Cherchneff (\cite{cherchneff}) shows that a departure from thermodynamical equilibrium due to shocks in the inner winds of M-type and carbon-rich  AGB stars results in higher abundances of $\rm{H_2O}$ in the inner parts of the CSEs. Previously, high amounts of $\rm{H_2O}$ have also been found in one C-type AGB star (IRC$+10216$/CW Leo; Melnick et al.~\cite{melnicketal01}; Hasegawa et al.~\cite{hasegawaetal}), indicating the need for a source of $\rm{H_2O}$ other than thermodynamical equilibrium chemistry. In the case of M-stars, the $\rm{H_2O}$ emission is expected to come from the warm, high-density innermost layers of the CSE, as supported by detailed radiative transfer modelling (e.g. Truong-Bach et al.~\cite{truongbachetal}; Maercker et al.~\cite{maerckeretal}). Hence, it is an important probe of the velocity structure in the region where the wind accelerates (Bains et al.~\cite{bainsetal}).

In Paper I we determined the circumstellar ortho-$\rm{H_2O}$ abundance in six M-type AGB stars, using the accelerated lambda iteration (ALI) technique for the radiative transfer, based on ISO LWS spectra between 43 and 197$\,\rm{\mu m}$. The ISO observations do not provide spectrally resolved line profiles. However, resolving the line profiles is essential, not only for the determination of accurate abundances, but particularly to gain information on the velocity field in the $\rm{H_2O}$ line-emitting region. In this paper we present the results of modelling the line profile of the ground-state ortho-$\rm{H_2O}$ line at 557 GHz, 1$_{10}$--1$_{01}$, observed with Odin (Nordh et al.~\cite{nordhetal}), towards R~Cas, R~Dor and W~Hya. The models are based on the results from Paper I, but we now take special consideration to the observed line profile, and examine how this helps to further constrain the ortho-$\rm{H_2O}$ abundance distribution and the velocity structure in the inner CSE.

Modelling of $\rm{H_2O}$ is not without difficulties. The high optical depths and subthermal excitation of $\rm{H_2O}$ complicate the radiative transfer considerably. Collisional excitation of the ground state can be important in particular for high mass-loss rate objects (Paper I), but is rather uncertain due to uncertainties in the calculated collisional cross sections (Phillips et al.~\cite{phillipsetal1996}). The effect of cooling due to $\rm{H_2O}$ line emission in the inner CSE adds additional complications. Finally, pumping through the vibrationally excited states due to IR photons at $2.6\,\mu$m and $6\,\mu$m from the central star and circumstellar dust needs to be taken into account.

In Paper I we showed the ability of our ALI code to handle the optical depths and subthermal excitation of the $\rm{H_2O}$ lines. We further estimated the effects of collisions and excitation through the first excited vibrational state of the bending mode ($\nu_2=1$, at $6.3\,\mu$m) on the models, and made a crude estimate of line cooling by $\rm{H_2O}$. The first excited vibrational states of the symmetric and asymmetric stretching modes ($\nu_1=1$ and $\nu_3=1$, corresponding to excitation at  $2.7\,\mu$m and $2.6\,\mu$m, respectively) may be particularly important for low mass-loss-rate objects. The high absorption in the stellar photosphere due to $\rm{H_2O}$ at these wavelengths makes the proper treatment of radiation at these wavelenghts especially complicated. We present here a first order estimate of how this affects the excitation.

We give a brief introduction to the Odin and ISO observations and basic information on the three objects in Section~\ref{observations}, and an overview of the modelling procedure in Section~\ref{CSE_mod}. In Sect.~\ref{results} we present the results of the modelling based on both Odin and ISO data, and discuss their effects on the predictions for the upcoming Herschel/HIFI mission. In Sect.~\ref{discussion} we discuss the results and to what degree the resolved line profiles can put constraints on the size of the $\rm{H_2O}$ line-emitting region and the velocity structure. Our conclusion are presented in section~\ref{conclusions}. 

\section{Observations}
\label{observations}

\subsection{Odin data}
\label{odindata}

\begin{figure*}
   \centering
   \includegraphics[width=18cm]{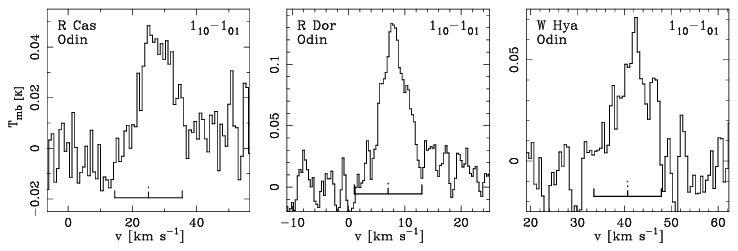}
      \caption{Odin observations of the ground-state transition of ortho-$\rm{H_2O}$, ($1_{10}-1_{01}$), at 557 GHz towards R~Cas, R~Dor, and W~Hya. The stellar systemic velocity, $v\rm{_{LSR}}$, is indicated by the dot-dashed line, and the solid lines indicate $v\rm{_{LSR}} \pm$$v\rm{_{e}}$ ($v\rm{_{e}}$ is the terminal gas expansion velocity determined in the CO models). }
              \label{odinlines}
    \end{figure*}

The  $\rm{H_2O}  $(1$_{10}$--1$_{01}$) lines (at 557 GHz) observed by Odin towards R~Cas, R~Dor and W~Hya are presented in Fig.~\ref{odinlines}. The W~Hya data was first presented in Justtanont et al. (\cite{justtanontetal}). Table~\ref{odin_data} gives a summary of the Odin observations, including the number of orbits and the epoch of observations. The integrated intensities ($I\rm{_{obs}}$) are given in $T\rm{_{mb}}$ scale, i.e. they are corrected for the main-beam efficiency of 90\%. The absolute calibration uncertainty is 10\%. The observations were done using the position-switch mode and the autocorrelation spectrometer (Frisk et al.~\cite{frisketal}; Olberg et al.~\cite{olbergetal2003}). The internal hot load was used for flux calibration. The data was rebinned after a sinusoidal baseline was fitted and subtracted. The frequency resolution is 1 MHz, which corresponds to $\rm{0.54\,km\,s^{-1}}$ at 557 GHz.

\begin{table}
\caption{Summary of Odin observations.}
\label{odin_data}
\centering
\begin{tabular}{l c c c c c}
\hline\hline
Source	& Orbits	& $t\rm{_{int}}$	& $I\rm{_{obs}}$ & 	$T\rm{_{mb}}$ & Epoch\\
		&		& [hours]		& [$\rm{K\,km\,s^{-1}}$] &	[K]	& \\
\hline
R~Cas	& 123	& 60.8		& 0.5$\pm$0.1	& 0.04$\pm$0.01&	2006.9\\
R~Dor	& 118	& 36.7		& 0.6$\pm$0.1	& 0.13$\pm$0.02&	2006.3\\
		&		&			&			&		            &     2006.4\\
W~Hya	& 185	& 53.4		& 0.4$\pm$0.1	& 0.06$\pm$0.02&	2002.9\\
		&		&			&			&		            &     2003.6\\
\hline\hline
\end{tabular}
\end{table}

\subsection{ISO observations}
\label{ISOdata}

In Paper I the models were fitted to ISO LWS spectra in the wavelength range of $43-197\,\rm{\mu m}$. The integrated intensities of the ortho-$\rm{H_2O}$Ê emission lines visible in the spectra were measured by fitting a Gaussian to the baseline-subtracted data after correcting the spectra for cosmic rays. The dominant error in the measured intensities is due to the subtraction of a baseline, and it is of the order of $\sim30\%$. For a more detailed description of the ISO observations, the data treatment, and the integrated intensities for our three objects, see Paper I.

\subsection{Observed sources}
\label{sources}

Our sources were studied already in Paper I, and we adopt the same stellar and circumstellar parameters in this paper. A brief summary of the source characteristics is given below.

\subsubsection{R~Cassiopeiae}
\label{rcas}

R~Cas is an M-type Mira variable with a period of 431 days (Kholopov et al.~\cite{kholopovetal1999}). A period-luminosity relationship based on bolometric magnitudes (Feast et al.~\cite{feastetal}) gives a luminosity of $8725\,L_{\odot}$. From this and the modelling of the spectral energy distribution (SED), the stellar distance is estimated to be 172 pc. The mass-loss rate, determined through modelling of CO radio rotational lines in Paper I, is $\rm{9\times10^{-7}\,M_{\odot}\,yr^{-1}}$. The gas expansion velocity of the CSE, derived in the CO emission line models, is $\rm{10.5\,km\,s^{-1}}$. Based on the ISO observations, the ortho-$\rm{H_2O}$ fractional abundance is estimated to be $\rm{3.5\times10^{-4}}$ (relative to $\rm{H_2}$; hereafter referred to as the abundance). See Sect.~\ref{dust_CO_mod} and Paper I for details on the modelling procedures.

\subsubsection{R~Doradus}
\label{rdor}

R~Dor is a semi-regular M-type AGB star with a period of 338 days (Kholopov et al.~\cite{kholopovetal1999}). The revised Hipparcos distance is 59~pc (Knapp et al.~\cite{knappetal}). The CO emission-line modelling in Paper I gives a mass-loss rate of $\rm{2\times10^{-7}\,M_{\odot}\,yr^{-1}}$, and a gas expansion velocity of $\rm{6.0\,km\,s^{-1}}$. The ortho-$\rm{H_2O}$ abundance, based on the modelling of the ISO data (Paper I), is $\rm{3\times10^{-4}}$. 

\subsubsection{W~Hydrae}
\label{whya}

W~Hya is a semi-regular M-type AGB star with a period of 382 days (Kholopov et al.~\cite{kholopovetal1999}), located at a distance of 78~pc (revised Hipparcos distance, Knapp et al.~\cite{knappetal}). The mass-loss rate is $\rm{1\times10^{-7}\,M_{\odot}\,yr^{-1}}$ (see Paper I), and the terminal gas expansion velocity is $\rm{7.2\,km\,s^{-1}}$. The ortho-$\rm{H_2O}$ abundance determined in Paper I is $\rm{15\times10^{-4}}$. Previous estimates of the $\rm{H_2O}$ abundance of this star, using various modelling techniques, agree with our estimate (e.g. Barlow et al.~\cite{barlowetal}; Zubko \& Elitzur~\cite{zubkoelitzur}; Justtanont et al.~\cite{justtanontetal}).

\section{The circumstellar emission modelling}
\label{CSE_mod}

The modelling of the circumstellar line emission was done in the same way as in Paper I, using radiative transfer codes based on the Monte Carlo and ALI methods to model lines from CO and ortho-$\rm{H_2O}$, respectively. Both codes include the radiation contribution from the circumstellar dust, which was modelled using the continuum radiative transfer code Dusty (Ivezi\'c et al.~\cite{ivezicetal}). The same `standard circumstellar model' as in Paper I is adopted, with constant mass-loss rate and expansion velocity. The inner radius of the CSE is set to the dust condensation radius determined in Dusty. 

\subsection{Dust and CO line emission modelling}
\label{dust_CO_mod}

The ortho-$\rm{H_2O}$ line models in this paper are based on the results of the dust and CO line emission modelling in Paper I. For model details and modelling procedure, see Paper I and references therein. The radiation from the dust is determined using Dusty to fit the SEDs, constructed using 2MASS and IRAS fluxes. The mass-loss rate (\emph{\.M}), gas expansion velocity ($v\rm{_{e}}$), and radial kinetic temperature profile of the CSE are determined through modelling of CO emission lines using the Monte Carlo method [presented in detail in Sch\"oier \& Olofsson~\cite{schoierolofsson01} and Olofsson et al.~\cite{olofssonetal02}]. Only the integrated intensities were fitted. In the three cases discussed here, there is a large difference between the predicted CO line profiles and the observed ones. The models consistently produce double-peaked line profiles, while the observations show single-peaked ones. That is, in the models the CO line emission is spatially resolved. In the case of W~Hya, Justtanont et al. (\cite{justtanontetal}) managed to obtain a better fit to the CO line profiles by reducing the envelope size and increasing the mass-loss rate. However, a similar approach does not work for R~Cas and R~Dor, indicating deviations from our standard circumstellar model (although the distance uncertainty also has an effect on the line profile). For the moment we do not have enough observational constraints to warrant a better mass-loss rate determination. The mass-loss-rate uncertainty contributes to the uncertainty in the absolute values of the derived abundances. However, the sensitivity of the 
$\rm{H_2O}$ line profiles to the different adopted parameters is largely unaffected by this mass-loss-rate uncertainty.

\subsection{$H_2O$ line emission modelling}
\label{H2O_mod}

\begin{table*}
\caption{Results of the best-fit models from paper I ($\rm{BF_{ISO}}$) and the best-fit models taking the Odin line profile into account ($\rm{BF_{Odin+ISO}}$) for R~Cas, R~Dor and W~Hya. Numbers within brackets give the best-fit values when excitation through the $\rm{\nu_3=1}$ state is included.}
\label{odin_iso}
\resizebox{\hsize}{!}{
\centering
\begin{tabular}{l c c c c c c c c c c c c c c c}
\hline\hline
Source	&  & &\multicolumn{6}{c}{$\rm{BF_{ISO}}$} & &\multicolumn{6}{c}{$\rm{BF_{Odin+ISO}}$} \\
 & \emph{\.M}& &$v\rm{_{e}}^a$  & $r\rm{_{e}}$ & $f\rm{_{0}}$ & $\delta\rm{_{Odin}}$ & $\delta\rm{_{ISO}}^b$ & $\chi_{red}^2$ & & $v\rm{_{e}}^a$ &$r\rm{_{e}}$ & $f\rm{_{0}}$ & $\delta\rm{_{Odin}}$ & $\delta\rm{_{ISO}}^b$ & $\chi_{red}^2$ \\
  &[$\rm{10^{-7}\,M_{\odot}\,yr^{-1}}$] & &[$\rm{km\,s^{-1}}$]  & [$\rm{10^{15}\,cm}$]  & [$10^{-4}$]& \% & \% & & &  [$\rm{km\,s^{-1}}$] &[$\rm{10^{15}\,cm}$]  & [$10^{-4}$]& \% & \% & \\
\hline
R Cas	&9& & 10.5			& 3.0&  \phantom{1}3.5		&  \phantom{1}+3	& --5	& 0.3		& & 8.5 & 3.0	&  \phantom{1}3.5	&  \phantom{1}1.2	& --3.6	& 0.3 \\
 & & & & & & & & & & & (3.0) & \phantom{1}(3.0) & \phantom{1}(6.8) & (--1.6) & (0.3) \\
R Dor	&2& & \phantom{1}6.0  		& 1.1	&   \phantom{1}3.0	& +46			& --1 & 0.6		& & 4.0 & 1.1	&  \phantom{1}1.5	&  \phantom{1}4.5	& 13.0	& 0.6\\ 
 & & & & & & & & & & & (1.3) & \phantom{1}(0.6) & \phantom{1}(--3.7) & (--5.8) & (0.6) \\
W Hya	&1& &  \phantom{1}7.2		& 2.2	&  15.0			& +67			& --1	& 0.5		& & 5.4 & 1.5	& 15.0			& --0.3	 		& --7.7 & 0.5\\
 & & & & & & & & & & & (2.2) & \phantom{1}(2.0) & \phantom{1}(--3.0) & (--7.8) & (0.4) \\
\hline
\hline
\end{tabular}}
\begin{list}{}{}
\item[$^{\rm{a}}$] $v\rm{_{e}}$ is determined by the CO line modelling in the $\rm{BF_{ISO}}$ models, and by the Odin line profile in the $\rm{BF_{Odin+ISO}}$ models.
\item[$^{\rm{b}}$] The difference $\delta$ between model and observed intensities is  an average difference of all lines for the ISO lines.
\end{list}
\end{table*}

We used the same ALI method as in Paper I in order to model the ortho-$\rm{H_2O}$ emission lines (model details and modelling procedure are described in Paper I). The molecular data for the ground state of ortho-$\rm{H_2O}$ was taken from the Leiden Atomic and Molecular Database\footnote{available at http://www.strw.leidenuniv.nl/$\sim$moldata} (LAMDA) (Sch\"oier et al.~\cite{schoieretal05}). The data for the vibrationally excited lines was taken from the high-resolution transmission (HITRAN) molecular absorption database (Rothman et al.~\cite{HITRAN}). The 45 lowest levels in the ground state of ortho-$\rm{H_2O}$ and of the first excited vibrational state of the bending mode ($\rm{\nu_2}=1$) are included, and the effect of excitation to the first excited vibrational state of the asymmetric stretching mode ($\rm{\nu_3}=1$) is tested (see Sect.~\ref{nu3}). Since the Einstein A-coefficients of the symmetric stretching mode ($\nu_1=1-0$) are an order of magnitude lower than for the asymmetric stretching, this state is not likely to affect the models and will not be considered here. Collisional rates in the ground state are taken from collisions between $\rm{He}$ and $\rm{H_2O}$, corrected by a factor of 1.4 to account for collisions with $\rm{H_2}$ (Green et al.~\cite{greenetal}). Collisions within the $\rm{\nu_2}=1$ state are taken to be the same as in the ground state, while collisions between the two vibrational states are scaled by a factor of 0.01. This scaling factor lies between the ones used by Menten et al.~(\cite{mentenetal}) and Josselin et al.~(\cite{josselin}) (0.02 and 0.005, respectively). The dust optical depth and temperature profile as determined by Dusty are used. The $\rm{H_2O}$ abundance distribution is assumed to be a Gaussian (centred on the star) defined by the e-folding radius $r\rm{_e}$ (where the abundance has decreased to 37\% of the initial value, $f\rm{_0}$):

\begin{equation}
\label{distribution}
f(r)=f_0\,\exp \left( -\left( \frac{r}{r_{\mathrm e}} \right)^2 \right)
\end{equation}

The mass-loss rate, gas expansion velocity, and kinetic temperature profile are included from the CO line emission modelling.  An estimate of $r\rm{_e}$ is given by the $\rm{H_2O}$ photodissociation calculations of Netzer \& Knapp~(\cite{netzerknapp}). We present `best-fit' models where the ortho-$\rm{H_2O}$ Êabundance ($f\rm{_0}$) and the size of the abundance distribution (in terms of the e-folding radius $r\rm{_e}$) were treated as free parameters, making it possible to set constraints on the photodissociation in the outer envelope (by constraining the size of the envelope), and the chemistry in the innermost envelope (by constraining the initial value of ortho-$\rm{H_2O}$). The best model fits are determined by fitting the model line fluxes to observed fluxes using the $\chi^2$ statistics. Paper I gives a detailed description of the numerical implementation of the ALI technique, as well as of the dependence of the model results on adopted numerical and physical parameters of the CSEs.

Basic results from the dust, CO line, and ortho-$\rm{H_2O}$ line emission modelling are given in Table~\ref{odin_iso}, including the reduced $\chi^2$ values of the models, $\chi^2_{red}$ The error in the determined abundances is estimated to be 50\% within the circumstellar model, and within a factor of $\sim5$ in absolute uncertainty.

\section{Results}
\label{results}

While Paper I concerns the determination of the ortho-$\rm{H_2O}$ abundance based on the spectrally unresolved ISO LWS lines, we here model the line profile of the ortho-$\rm{H_2O}$ 557 GHz line observed with the Odin satellite in order to further constrain the models and derive new results on e.g. abundance, molecular envelope size, and gas expansion velocity (in the region where the 557 GHz line is emitted). The effect of these new results on the line profiles for lines within the range of the upcoming Herschel/HIFI mission is also examined.

\subsection{The 557 GHz Odin line predicted by the ISO models}
\label{Odin_basic}

\begin{figure*}
   \centering
   \includegraphics[width=18cm]{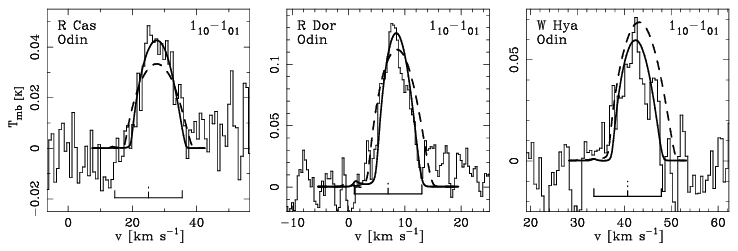}
      \caption{The ortho-$\rm{H_2O}$ ($1_{10}-1_{01}$) line at 557 GHz observed with Odin towards R~Cas, R~Dor, and W~Hya, overlaid by the $\rm{BF_{ISO}}$ model from Paper I (dashed line), and the $\rm{BF_{Odin+ISO}}$ model (solid line). The stellar systemic velocity, $v\rm{_{LSR}}$, is indicated by the dot-dashed line, and the solid lines indicate $v\rm{_{LSR}} \pm$$v\rm{_{e}}$ ($v\rm{_{e}}$ is the gas terminal expansion velocity determined in the CO model). }
              \label{odin_obs}
    \end{figure*}

Figure~\ref{odin_obs} shows the $\rm{H_2O}$ line at 557 GHz observed towards R~Cas, R~Dor and W~Hya, overlaid by the model lines based on the $\rm{BF_{ISO}}$ models from Paper I (dashed line). The differences between model and observed fluxes for the Odin and ISO lines ($\delta\rm{_{Odin}}$ and $\delta\rm{_{ISO}}$, respectively) are given in \% in Table~\ref{odin_iso}. Most notably, the model lines are broader than the observed lines. This indicates that the 557 GHz line probes a region with a lower expansion velocity than the CO lines. In addition, the integrated flux of the Odin line is somewhat overestimated in all cases. Of importance here is also the fact that, for all three objects, the 557 GHz line is optically thick throughout the $\rm{H_2O}$ envelope, hence the emitting region of this line is limited by the envelope size, whereas the ISO line regions are excitation limited. This makes the 557 GHz line particularly sensitive to the outer radius of the $\rm{H_2O}$ envelope.

\subsection{Line shape dependence on CSE parameters}
\label{param_dep}

\begin{figure}
   \centering
   \includegraphics[width=5.72cm,angle=-90]{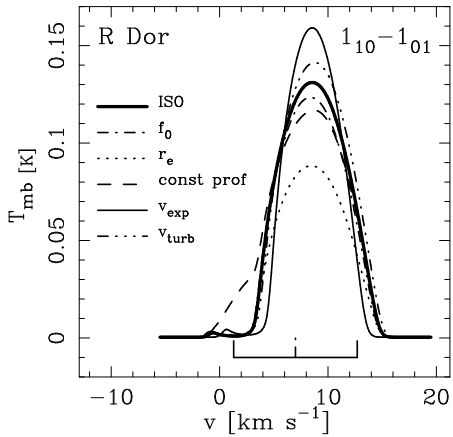}
      \caption{The $\rm{H_2O}$(1$_{10}$--1$_{01}$) line profile dependence on various parameters.  One parameter  is varied at a time (the varied parameter is indicated in the left part of the figure). ISO refers to the best-fit model based on ISO data only from Paper I. The label \emph{const prof} refers to the model with a step function abundance distribution. The parameters are varied  according to Table~\ref{flux_change} (only the results of a reduction in abundance and radius is shown). The stellar systemic velocity $v\rm{_{LSR}}$ is indicated by the dot-dashed line, and the solid lines indicate $v\rm{_{LSR}} \pm$$v\rm{_{e}}$ ($v\rm{_{e}}$ is the terminal gas expansion velocity determined in the CO models).}
              \label{line_change}
    \end{figure}

Test models were run to determine the general effect of individual parameters on the $\rm{H_2O}$(1$_{10}$--1$_{01}$) line profile. The main parameters which may affect the line profile are the abundance $f\rm{_{0}}$, the size of the envelope $r\rm{_e}$, the shape of the abundance distribution, the turbulent velocity width $v\rm{_{turb}}$, and the expansion velocity $v\rm{_{e}}$ (which, at a constant mass-loss rate, effectively changes the density profile).  As an example, Fig.~\ref{line_change} shows the effects of these different parameters on the line profile of R~Dor. The difference in the line fluxes for the Odin and ISO lines compared to the $\rm{BF_{ISO}}$ model fluxes is given in Table~\ref{flux_change}. Other parameters (the mass-loss rate, kinetic temperature profile, collisional excitation efficiency, the dust and stellar radiation fields, and the excitation to the $\nu_2$ state) affect the integrated intensity of the 557 GHz line to the same degree as the ISO lines (tested in Paper I), while its line profile does not change significantly. 

In terms of the total flux, Table~\ref{flux_change} shows that the 557 GHz line is sensitive to the ortho-$\rm{H_2O}$ abundance and outer radius of the $\rm{H_2O}$ envelope (in particular compared to the sensitivity of the ISO lines to the outer radius). Whereas the integrated line flux is less sensitive to the remaining parameters, the shape of the line changes noticeably. In particular, the reduction in expansion velocity results in a narrower line profile and a significantly higher peak intensity ($+23\%$ in the case of R~Dor compared to the best-fit ISO model, see Fig.~\ref{line_change}). Changing the abundance distribution ($f$-distr.) to a step function (i.e. a constant abundance $f\rm{_0}$, which suddenly drops to zero at $r\rm{_e}$) results in less self-absorption in the blue wing of the line, and it leads to a decrease in the peak intensity ($-11\%$ for R~Dor). The reduced self-absorption in models with a step function abundance distribution is due to the sudden cut-off of the ortho-$\rm{H_2O}$ abundance at $r\rm{_e}$, removing $\rm{H_2O}$ present at further distances responsible for the increased self absorption in models using a Gaussian distribution. Increasing the turbulent velocity width increases the amount of self-absorption. As can be seen in Fig.~\ref{line_change}, the most significant change in the shape of the line profile occurs when changing the abundance distribution. 

Note that the models predict a significant amount of emission in the line wings at velocities outside the range $v\rm{_{LSR}} \pm$$v\rm{_{e}}$. Due to the relatively small size of the $\rm{H_2O}$ CSEs, lines with optical depths around or larger than unity in the outer parts of the CSEs are expected to have a significant contribution of the local line profile to the line wings. This means that the observed 557 GHz lines are expected to be wider than 2$v\rm{_{e}}$.

\begin{table}
\caption{The change in integrated flux compared to the best-fit models using ISO data only. Only one parameter is varied at a time.}
\label{flux_change}
\resizebox{\hsize}{!}{
\centering
\begin{tabular}{l r r r r r r r }
\hline\hline
Param. & change & \multicolumn{2}{c}{R~Cas} & \multicolumn{2}{c}{R~Dor} & \multicolumn{2}{c}{W~Hya}  \\
              &                & $\delta\rm{_{Odin}}$ & $\delta\rm{_{ISO}}$ & $\delta\rm{_{Odin}}$ & $\delta\rm{_{ISO}}$ & $\delta\rm{_{Odin}}$ & $\delta\rm{_{ISO}}$\\
\hline
$f\rm{_{0}}$				& +100\%	& +30\%	& +21\%	& +19\%	& +22\%	& --2\%	& +15\%	\\
						& --50\%	& --20\%	& --21\% 	& --6\% 	& --18\% 	& --11\%	& --12\% \\
$r\rm{_e}$				& +100\%	& +50\%	& +11\%	& +30\%	& +11\%	& +34\%	&0\%	\\
						& --50\%	& --35\%	& --14\%	& --32\% 	& --18\% 	& --34\%	& --18\% \\
$f$-distr. 					& step func.& +7\%	& +7\%	& +1\%	& +9\%	& -6\%	& +9\% \\
$v\rm{_{e}}$				& --20\%	& +6\%	& +9\%	& --4\% 	& +8\%	& --9\%	& +3\% \\
$v\rm{_{turb}}$				& +50\% 	& +2\%	& +1\%	& +7\%	& +9\%	& +7\%	& +4\%	\\
$\nu_3=1$				& incl.	& +10\%	& +17\%	& +2\%      & +35\%   & +65\%   &+24\% \\
\hline\hline
\end{tabular}}
\end{table}

\subsection{Odin and ISO data combined}
\label{odin_ISO}

  \begin{figure*}
   \centering
   \includegraphics[width=18cm]{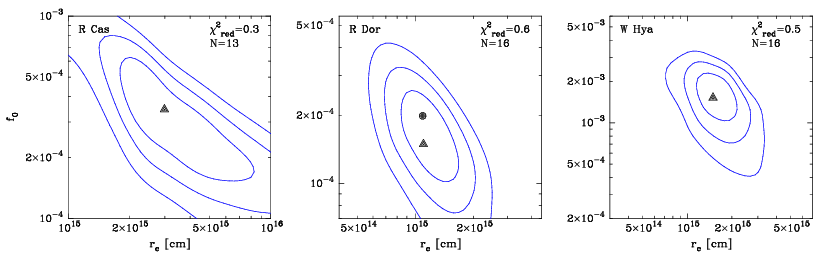}
   \caption{$\chi^2$ maps in ortho-$\rm{H_2O}$ abundance and radius of the $\rm{H_2O}$ envelope for R~Cas, R~Dor and W~Hya, where the modelling takes both the ISO and Odin lines into account. The triangles mark the best-fit models $\rm{BF_{Odin+ISO}}$. For R~Dor the dot marks the lowest $\chi^2$ value (which is obtained for R~Cas and W~Hya when the 557 GHz line profile is best fitted). $N$ gives the total number of lines fitted in the modelling.}
              \label{odin_chi2}
    \end{figure*}

In Paper I, the ortho-$\rm{H_2O}$ models were fitted using only the ISO LWS spectra, giving rather uncertain constraints on the ortho-$\rm{H_2O}$ abundance and, in particular, the size of the $\rm{H_2O}$ abundance distribution. The Odin line at 557 GHz is well excited throughout the $\rm{H_2O}$ envelope, and it is therefore likely to set a tighter constraint on especially the spatial extent of the $\rm{H_2O}$ envelope. The effect of the 557 GHz line is tested by fitting models including both the ISO and Odin integrated fluxes. The models are varied in ortho-$\rm{H_2O}$ abundance and outer radius $r\rm{_e}$, and a new $\chi^2$ analysis is done. 

The gas expansion velocity is adjusted to fit the line widths of the $\rm{H_2O}$(1$_{10}$--1$_{01}$) lines. This requires a reduction in $v\rm{_{e}}$, compared to that obtained from the CO models, by 19\%, 33\%, and 25\% for R~Cas, R~Dor and W~Hya, respectively. Although the data has only moderate signal-to-noise ratios, a lower expansion velocity not only produces a significantly better fit to the line shape, but also a better fit to the integrated flux of the Odin line and lower $\chi^2$ values in all three cases. Furthermore, observations of the para-$\rm{H_2O}$ line at 183~GHz in maser emission (Gonz\'alez-Alfonso et al.~\cite{gonzalez-alfonsoetal1998}) indicate expansion velocities for W~Hya and R~Cas of $\approx\,\rm{5\,km\,s^{-1}}$ and  $\approx\,\rm{7\,km\,s^{-1}}$, respectively. Thus, we are confident that the lower gas expansion velocities derived from the Odin lines are real.

Figure~\ref{odin_chi2} shows the resulting $\chi^2$ maps. Compared to the corresponding maps presented in Paper I (Fig. 3 in Paper I), it is obvious that the inclusion of the 557 GHz line sets much stronger constraints on the envelope size than models only including the ISO data. For R~Cas and W~Hya the models with the lowest $\chi^2$ values also give the best-fit to the 557 GHz line profile. A slight reduction of the abundance by 50\%, compared to the lowest $\chi^2$ model, gives an even better fit to the line profile for R~Dor. The best-fit models, $\rm{BF_{Odin+ISO}}$, are marked with triangles in Fig.~\ref{odin_chi2}, and their results are given in Table~\ref{odin_iso}. For R~Dor the model with the lowest $\chi^2$ value is marked by a dot in Fig.~\ref{odin_chi2}.

The resulting 557 GHz line profiles are shown in Fig.~\ref{odin_obs} as the solid lines. The difference between the model ISO and Odin fluxes and the observed fluxes is given in \% in Table~\ref{odin_iso} (for the ISO lines the average difference is given). In all cases the difference is well below the observational error. For R~Cas and R~Dor the radius $r\rm{_e}$ is the same as in the $\rm{BF_{ISO}}$ models - the inclusion of the 557 GHz line only leads to a tighter constraint on the envelope size. For W~Hya the inclusion of the 557 GHz line results in a new best-fit radius that is 32\% smaller, compared to the $\rm{BF_{ISO}}$ model.  The best-fit abundances do not change for R~Cas and W~Hya, while inclusion of the 557 GHz line results in a decrease of the abundance of 50\% for R~Dor.

\subsection{HIFI predictions revisited}
\label{HIFI}

  \begin{figure*}
   \centering
   \includegraphics[width=18cm]{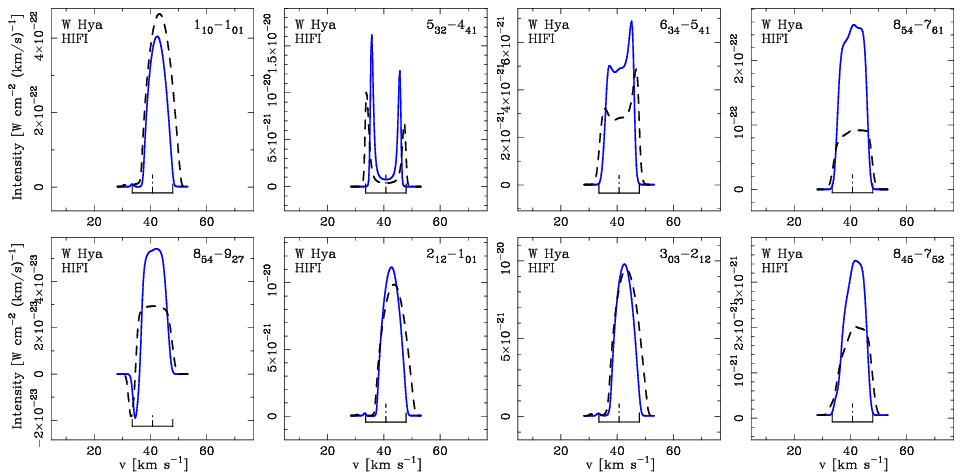}
   \caption{Eight ortho-H$_2$O lines that can be observed with Herschel/HIFI for W~Hya. The solid line shows $\rm{BF_{Odin+ISO}}$ models. The dashed line shows the $\rm{BF_{ISO}}$ models for comparison. The stellar systemic velocity $v\rm{_{LSR}}$ is indicated by the dot-dashed line, and the solid lines indicate $v\rm{_{LSR}} \pm$$v\rm{_{e}}$ ($v\rm{_{e}}$ is the terminal gas expansion velocity determined in the CO models).}
              \label{hifilines}
    \end{figure*}

Figure~\ref{hifilines} Êshows the difference in the HIFI line profiles for W~Hya between the $\rm{BF_{ISO}}$ models (dashed lines) and the $\rm{BF_{Odin+ISO}}$ models taking the 557 GHz line profile into consideration (solid lines). The decrease of the outer radius causes the HIFI lines to decrease somewhat in integrated intensity, whereas the reduced expansion velocity leads to an increased line intensity. These effects generally cancel each other in the optically thick lines. For optically thin lines (e.g. the two masering lines $5_{32}-4_{41}$ and $6_{34}-5_{41}$, and the lines $8_{54}-7_{61}$, $8_{54}-9_{27}$, and $8_{45}-7_{52}$) the reduction in expansion velocity results in an increase of the integrated and peak intensities of approximately a factor of two. Thus, the prediction in Paper I that these lines are observable (some of them easily) with Herschel/HIFI still holds. Note that the HIFI lines come from different parts of the CSE and different expansion velocities may apply to them. This will of course affect the line profile and results presented in Fig.~\ref{hifilines} should be regarded with some caution.

In terms of the line profile dependence on CSE parameters, the HIFI lines react similar to the Odin line in Fig.~\ref{line_change}, i.e. reduced self-absorbtion in step function abundance distribution models, narrower lines with reduced expansion velocity, and decreased peak and integrated intensities with a decrease in abundance and/or outer radius.

\subsection{Effects of excitation through the $\nu_3=1$ state}
\label{nu3}

In order to estimate the effect of excitation through the $\rm{\nu_3}=1$ state, we include the 43 lowest energy levels of this state in our model and repeat the $\chi^2$ analysis. The new best-fit results are presented in brackets in Table~\ref{odin_iso}. The change in integrated line fluxes when including the $\rm{\nu_3}$ state are presented in Table~\ref{flux_change}. The average change in integrated flux for the ISO lines is dominated by the $7_{43}-6_{43}$ transition at 116.78$\,\mu$m for R~Dor and W~Hya (the line changes by 170\% and 150\%, respectively). Excluding this line results in an average change in the ISO lines of 25\% and 15\% for R~Dor and W~Hya, respectively.

For all three objects the change in the $\rm{H_2O}$(1$_{10}$--1$_{01}$) line profile is very small when including the $\rm{\nu_3}=1$ state. A decrease in optical depth leads to less self absorption on the blue side and less broadening on the red side.

The Odin line in the W~Hya models is, perhaps somewhat surprisingly, affected more strongly than in the other two objects. This is due to the combination of a low mass-loss rate and the very high initial abundance in the best-fit model when only the $\rm{\nu_2}$ state is included. [The change in the integrated flux using an abundance one order of magnitude lower is +4\% for the Odin line and $\approx40\%$ on average for the ISO lines, more in line with the other objects. Also, increasing the mass-loss rate by a factor of two moderates the effect of the $\nu_3$\,=\,1 excitation.]

Inspecting the effect of excitation through the $\rm{\nu_3}$ state on the HIFI lines we find that the integrated flux increases on average by 25\%, 54\%, and 58\% for R~Cas, R~Dor, and W~Hya, respectively. However, a few lines change by considerably more ($\approx100-200\%$), while others are much less sensitive (with changes in integrated flux of only a few percent).

It is clear that a detailed sampling of the radiation from the stellar photosphere at $2.6\,\mu$m is required to properly include the first excited vibrational state of the asymmetric stretching mode ($\rm{\nu_3=1}$) in the models. Our models approximate the stellar radiation by a simple black-body, likely overestimating the flux at $2.6\,\mu$m (see Sect.~\ref{details}).

\section{Discussion}
\label{discussion}

\subsection{Additional constraints}
\label{addconst}

The Odin line profiles can be fitted simultaneously with the ISO lines without requiring a change in the simple approximations made in the `standard circumstellar model'. However, since we are only fitting one resolved line, the assumptions in the circumstellar model (e.g. constant $v\rm{_{e}}$, constant $\emph{\.M}$) may be correct only within the region from which the line is coming. When Herschel/HIFI data becomes available and the line profiles of a series of lines have to be fitted, the situation may change. As the lines probe different regions in the CSE, fitting different lines may require a gradient in the velocity distribution, i.e. an acceleration zone, and a different abundance distribution (see Sect.~\ref{velconst}). 

\subsection{The size of the $H_2O$ envelope}
\label{h2oregion}

\begin{figure*}[ht]
   \centering
   \includegraphics[width=18cm]{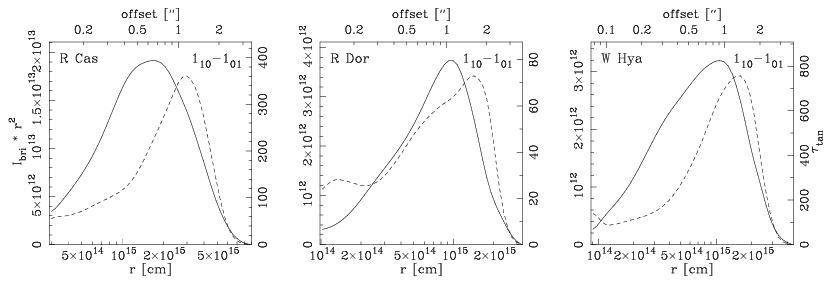}
      \caption{The tangential optical depth (dashed lines) in the ortho-$\rm{H_2O}$ ($1_{10}-1_{01}$) line at 557 GHz and the brightness distribution (solid lines) in the $\rm{BF_{Odin+ISO}}$ models for R~Cas, R~Dor, and W~Hya.}
              \label{odin_tau}
    \end{figure*}
    
For all three objects, the Odin line at 557 GHz puts an additional constraint on the $\rm{H_2O}$ envelope size. Figure~\ref{odin_tau} shows the tangential optical depth of the Odin line for R~Cas, R~Dor and W~Hya. The line is excited throughout the envelope, until the ortho-$\rm{H_2O}$ abundance drops at the e-folding radius $r\rm{_e}$. In Paper I, the $\rm{BF_{ISO}}$ model radius was found to be within 1$\sigma$ of the theoretical radius given by Netzer \& Knapp (\cite{netzerknapp}). The latter is obtained from models of $\rm{H_2O}$ photodissociation and it is defined as the distance where the abundance of OH is highest. Although the 557 GHz data requires a reduction in the outer radius, the new best-fit radii based on the combined Odin and ISO data are still within the 1$\sigma$ contours presented in Paper I, though with much tighter constraints, confirming the photodissociation models of  Netzer \& Knapp (\cite{netzerknapp}). Observations of the 1665 and 1667 MHz OH masers in the CSE around R~Cas (Chapman et al.~\cite{chapmanetal}) and W~Hya (Szymczak et al.~\cite{szymczaketal}) show OH masering shells at $(1.3-2.6)\times10^{15}$cm and $\rm{(0.8 - 1.3)\times10^{15}\,cm}$ from the central star, respectively. This is consistent with the smaller radii of the $\rm{H_2O}$ envelopes determined for R~Cas and W~Hya in this paper.

\subsection{The abundance of $\rm{H_2O}$}
\label{abundance}

The derived total (para $+$ ortho) $\rm{H_2O}$ abundances (assuming an ortho-to-para ratio of 1) in Paper I are $\sim\,$6$\,\times10^{-4}$.  The thermal equilibrium abundance of $\rm{H_2O}$ is approximately 1$-$3$\,\times\,10^{-4}$ (Mamon et al.~\cite{mamonetal87}; Cherchneff~\cite{cherchneff}), while the upper limit given by the cosmic abundances of carbon and oxygen is no more than $1\times10^{-3}$, assuming that all carbon goes into CO and the remaining oxygen into $\rm{H_2O}$ (Anders \& Grevesse~\cite{andersgrevesse}). Models of departures from thermodynamical equilibrium chemistry due to changes in density and temperature as a result of shocks in the inner parts of the winds predict total (para $+$ ortho) $\rm{H_2O}$ abundances on the order of $\approx4\,\times\,10^{-4}$ (Cherchneff~\cite{cherchneff}). This likely explains at least part of the observed high abundances of $\rm{H_2O}$. Other processes which may explain an increased abundance of $\rm{H_2O}$ are the evaporation of icy bodies (Justtanont et al.~\cite{justtanontetal}) or grain surface reactions (Willacy~\cite{willacy}).
However, the inclusion of the Odin line and the $\rm{\nu_3=1}$ mode in the excitation analysis result in lower abundances.  Considering the current absolute uncertainties (at least on the order of a factor of 5), the results cannot conclusively confirm or rule out the various competing scenarios, from thermodynamical  equilibrium chemistry to the evaporation of icy bodies.

\subsection{The acceleration zone}
\label{velconst}

In dust-driven winds, the dust accelerates and initiates the mass loss relatively close to the star. Depending on the size of the $\rm{H_2O}$ line-emitting region, observations of water vapour lines may probe the inner structure of the CSE and give information on the velocity field. A number of observations indicate that the $\rm{H_2O}$ line-emitting region coincides with the acceleration zone in M-type AGB stars. Imaging in the 22 GHz $\rm{H_2O}$ maser line, indicates that the terminal velocity (i.e. the final gas expansion velocity) is reached at relatively large distances from the central stars (Bains et al.~\cite{bainsetal}). For R~Cas, the analysis of $\rm{H_2O}$ masers by Colomer et al. (\cite{colomeretal2000}) indicates an expansion velocity of only $3.0\,\rm{km\,s^{-1}}$ in the inner regions of the envelope ($\approx\,3.4\times10^{14}$cm; the gas terminal expansion velocity estimated from CO data is $\rm{10.5\,km\,s^{-1}}$). Similarily, models fitted to OH, $\rm{H_2O}$ and SiO masers in W~Hya indicate that the terminal expansion velocity (estimated from CO data to be $\rm{7.2\,km\,s^{-1}}$) is reached at distances $\rm{\gtrsim2\times10^{15}\,cm}$ (Szymczak et al.~\cite{szymczaketal}). The $\rm{H_2O}$ maser is located in a ring-like structure with a radius of $\rm{1.8\times10^{14}\,cm}$ expanding with a velocity of $\rm{3.6\,km\,s^{-1}}$ (Reid \& Menten~\cite{reidmenten}), and the OH masers lie at $\rm{8.2\times10^{14}\,cm}$ and $\rm{1.3\times10^{15}\,cm}$ with expansion velocities of $\rm{4.2\,km\,s^{-1}}$ and $\rm{5.2\,km\,s^{-1}}$, respectively (Szymczak et al.~\cite{szymczaketal}). The SiO maser has its origin at a distance of $\rm{5.8\times10^{13}\,cm}$ (Miyoshi et al.~\cite{miyoshietal94}), and the lower limit of the expansion velocity is estimated to be about $\rm{2\,km\,s^{-1}}$ (Szymczak et al.~\cite{szymczaketal}). Likewise, in order to model high angular resolution observations of SiO in the CSE around R~Dor, Sch\"oier et al. (\cite{schoieretal2004}) determined a best-fit model to the data using a two component density distribution (with a higher SiO abundance within $1.0\times10^{15}$cm) and a lower expansion velocity in the inner regions of the CSE of $4.9\,\rm{km\,s^{-1}}$ (the gas terminal expansion velocity estimated from CO data is $\rm{6.0\,km\,s^{-1}}$). 

The fits to the Odin line profile indicate that the $\rm{H_2O}$ line-emitting region covers an area in which the expansion velocity has not yet reached its terminal value. As can be seen in Fig.~\ref{odin_tau}, the dominant part of the $\rm{H_2O}$ 557 GHz line emission comes from a region at a radius of about 10$^{15}$\,cm for all three sources. For R~Cas, R~Dor, and W~Hya, good fits to the 557 GHz line profiles require expansion velocities of only 8.5, 4.0, and $5.4\,\rm{km\,s^{-1}}$, respectively. This means a 19, 33, and 25\%, respectively, decrease from the terminal gas expansion velocities as measured in the CO radio lines. Thus, the 557 GHz line data is consistent with an acceleration zone extending beyond 10$^{15}$\,cm for these, relatively, low-mass-loss-rate objects. 

\begin{figure}
   \centering
   \includegraphics[width=5cm,angle=270]{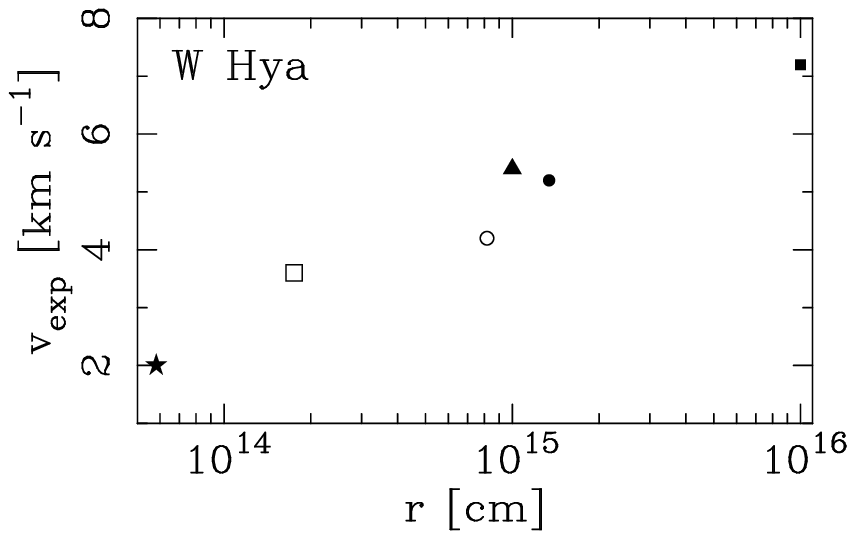}
      \caption{Expansion velocity data for W~Hya obtained from various probes. The data points are from SiO masers (star), $\rm{H_2O}$ masers (open box), OH masers at 1665 MHz (open circle) and 1667 MHz (dot), the fit to the Odin 557 GHz line data (triangle), and the CO expansion velocity (filled box). See text for more details.}
              \label{whya_vprof}
    \end{figure}

\begin{figure*}
   \centering
   \includegraphics[width=18cm]{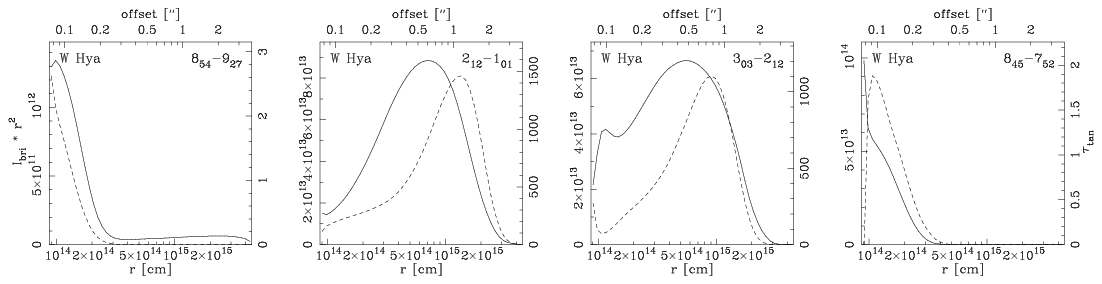}
   \caption{The tangential optical depths (dashed lines) in the $\rm{BF_{Odin+ISO}}$ models and their brightness distributions (solid lines) for four lines that can be observed by Herschel/HIFI towards W~Hya.}
              \label{HIFI_tau}
    \end{figure*}

Fig.~\ref{whya_vprof} shows the estimated positions of the masers and the corresponding gas expansion velocities for W~Hya. The expansion velocity required to fit the Odin 557 GHz line profile has been placed at a radius of $\rm{1\times10^{15}\,cm}$ (where the brightness distribution peaks), and the terminal gas expansion velocity determined by the CO emission lines is placed at $\rm{1.0\times10^{16}\,cm}$ (the estimated CO photodissociation radius, Olofsson et al.~\cite{olofssonetal02}). The measurements show that the wind has not reached its terminal value in the region where the 557 GHz line originates. Figure~\ref{HIFI_tau} shows the tangential optical depth distributions of four of the HIFI lines shown in Fig.~\ref{hifilines} for W~Hya. Two of the lines probe a region similar to that of the 557 GHz line, whereas the other two appear to be emitted very close to the star. The HIFI observations of $\rm{H_2O}$ will likely probe different regions of the CSE, and in particular may probe regions with different expansion velocities. Fitting the individual line profiles of the HIFI observations should therefore give important information on the velocity structure of the CSE.

\subsection{The importance of detailed models}
\label{details}

Inclusion of excitation to the first excited vibrational state of the asymmetric stretching mode ($\rm{\nu_3}=1$) changes the integrated fluxes of the modeled lines significantly, in particular for the low-mass-loss-rate objects R~Dor and W~Hya. On the contrary, Gonz\'alez-Alfonso et al.~(\cite{gonzalez-alfonsoetal2007}), when modelling the $\rm{1_{10}-1_{01}}$ transition observed with the Submillimeter Wave Astronomy Satellite (SWAS) in the carbon star IRC+10216, concluded that the $\rm{1_{10}-1_{01}}$ transition is not significantly affected by pumping through the $\nu_3$\,=\,1 state, as the pumping of the line occurs mainly through the $\nu_2$\,=\,1 state in this object. However, the abundance of o-$\rm{H_2O}$ in their models is low, and IRC+10216 has a high mass-loss rate. We find the same trend with mass-loss rate and abundance in our models of the $\rm{1_{10}-1_{01}}$ emission. Models of photon dominated regions show that IR pumping affects higher lying transitions more strongly in accordance with our results for lower abundances (Gonzalez Garcia et al.~\cite{gonzalezgarciaetal2008}).

It must be emphasized here that our models approximate the radiation from the central star with a blackbody, and therefore likely overestimate the flux at $2.6\,\mu$m as photospheric absorption by $\rm{H_2O}$ is not taken into account. This is confirmed by low-resolution SWS spectra from ISO for R~Dor (Ryde \& Eriksson~\cite{rydeco2002}) and W~Hya (Justtanont et al.~\cite{justtanont2004}). It is therefore difficult to accurately include the effect of the $\rm{\nu_3}$\,=\,1 state without detailed modelling of the photosphere. The $\rm{H_2O}$ abundance for the low-mass-loss-rate objects will likely lie between the abundances determined when including both the $\rm{\nu_2}$\,=\,1 and $\rm{\nu_3}$\,=\,1 states and when including only the $\rm{\nu_2}$\,=\,1 state (assuming a blackbody radiation field). This demonstrates how important detailed models will be in order to determine accurate $\rm{H_2O}$ abundances from Herschel/HIFI data.

\section{Conclusions}
\label{conclusions}

We have modelled the ortho-$\rm{H_2O}$($1_{10}-1_{01}$) line at 557 GHz observed with the Odin satellite for the M-type AGB stars R~Cas, R~Dor and W~Hya. The models are based on previous models of ortho-$\rm{H_2O}$ lines observed with ISO (Paper I). The Odin line profile is resolved in contrast to the ISO lines, and therefore gives additional constraints to the circumstellar model. Models that fit the integrated intensities of both the ISO and Odin lines result in tighter constraints on the size of the $\rm{H_2O}$ envelope than models based on ISO observations only. The resulting $\rm{H_2O}$ envelopes are smaller than derived from the ISO data alone. The envelope sizes given by models of $\rm{H_2O}$ photodissociation lie within the errors of our estimates, confirming the photodissociation chemistry in the outer envelope. Additionally, expansion velocities lower than those estimated from CO data are required to fit the 557 GHz line profile, hence indicating that observations of water vapour probe the acceleration zone in the CSE. The estimated sizes of the $\rm{H_2O}$ envelopes and the expansion velocities are confirmed by observations of $\rm{H_2O}$ and OH maser emission. 

The $\rm{H_2O}$ abundances from Paper I are confirmed here and, taken at face value, cannot be explained by thermodynamical equilibrium chemistry. However, the effect of excitation through the first excited vibrational state of the asymmetric stretching mode ($\rm{\nu_3=1}$) is tested (assuming a stellar blackbody field), and it leads to significant decreases in abundances for the low-mass-loss-rate objects. This shows the difficulty in accurately modelling circumstellar $\rm{H_2O}$ emission lines. Careful and detailed modelling of the stellar radiation field is necessary in order to determine correct $\rm{H_2O}$ abundances, and hence identify the origin of water vapour in the CSEs of AGB stars, and to determine important properties of the inner CSEs. This will be particularly important for the upcoming Herschel/HIFI mission.

The HIFI observations will help to set constraints on the $\rm{H_2O}$ abundance distribution as well as the kinematics of the CSE. HIFI/Herschel will observe $\rm{H_2O}$ Êemission lines in the range 557 to 1885 GHz, covering a range of excitation conditions and optical depths. While the lower lying transitions (like the 557 GHz line) will constrain the size of the emitting region (as they are sensitive to the outer radius), higher lying transitions will set constraints on the velocity structure close to the star. As these two parameters are constrained by fitting the line profile for a range of emission lines, the abundance estimates made in Paper I and here will likely be improved significantly. 

\begin{acknowledgements}
The authors would like to thank the anonymous referee for comments that helped to improve the quality of the analysis. The authors acknowledge the financial support from the Swedish Research Council and the Swedish National Space Board.
\end{acknowledgements}

\end{document}